\documentclass[useAMS,usenatbib]{mn2e}
\usepackage{amsmath,amsfonts,amssymb}

\RequirePackage[dvips]{graphicx}

\usepackage{color}
\def\aap{\emph{A.\& A.}}

\def\apjs{\emph{ApJ. Supp.}}
\def\apjl{\emph{ApJ. Lett.}}

\title[Infrahumps detected in V1504 Cygni]{Infrahumps detected in Kepler light curve of V1504 Cygni}

\author[Coyne et al.]{R. Coyne\thanks{E-mail:
robcoyne@gwu.edu (RC)}, A. Shenoy, G.A. MacLachlan, T.R. Lewis, K.S. Dhuga, 
\newauthor A. Eskandarian, B.E. Cobb, L.C. Maximon, and W.C. Parke\\
Department of Physics, The George Washington University, Washington, D.C. 20052, USA.\\
}

\begin{document}

\maketitle

\label{firstpage}

\begin{abstract}
We present a power spectral density analysis of the short cadence 
Kepler data for the cataclysmic variable V1504 Cygni. 
We identify three distinct periods: the orbital period ($1.668\pm0.006$ hours), the 
superhump period ($1.732\pm0.010$ hours), and the infrahump period ($1.633\pm0.005$ hours). 
The results are consistent with those predicted by the period excess-deficit relation.

\end{abstract}


\section{Introduction}\label{Introduction}
NASA's \emph{Kepler} mission has been providing high 
time-resolution optical photometry 
since its launch on March 6, 2009 \citep{koch:2010}. Designed to 
survey a fixed 105 square degree field of 
view, it monitors the brightness of over 100,000 stars primarily for 
the purpose of detecting and categorizing exoplanets. However, due to 
the extremely high quality of the instrument, it has proven itself a boon 
for other astronomical communities as well, including those interested in 
the study of Cataclysmic Variables (CVs).

CVs are classified based on their outburst properties, with the most 
active category being dwarf novae (DN). These systems are marked by 
quasi-periodic outbursts of varying strength, typically increasing in 
brightness by a few magnitudes. Dwarf novae are further separated into 
subtypes based on the manner in which these outbursts present 
themselves. \textit{SU UMa}-type DN are characterized by the presence 
of occasional superoutbursts, dramatic eruptions several magnitudes 
brighter than normal outbursts. DN that do not exhibit such behavior 
are classified as \textit{U Gem}-type. 

Whether a system exhibits \textit{SU UMa} or \textit{U Gem} 
characteristics is dependent on the orbital dynamics of the system, 
and can be characterized by its mass ratio ($q=M_2/M_1$) and its 
orbital period (see Eq. \ref{eq:smith-dhillon-rel}). 
Each group tends to lie within separate distributions 
on either side of a critical mass ratio 
($q_\mathrm{crit}\approx0.33$). \textit{U Gem}-type DN have long 
periods corresponding to mass ratios greater than the critical 
value ($q>q_\mathrm{crit}$) whereas \textit{SU UMa}-type DN have 
shorter periods corresponding to mass ratios less 
than the critical value ($q<q_\mathrm{crit}$) \citep{hellier:2001}.

The \textit{Kepler} field of view includes 10 CVs. 
Of the \textit{SU UMa}-type, only V344 Lyra has been the subject 
of extensive study \citep{wood:2011}, and to date no equivalent treatment 
has been undertaken for V1504 Cygni, which is the object of interest in 
this paper. Like V344 Lyra, V1504 Cygni is a \textit{SU UMa}-type 
dwarf novae, and as such features superoutbursts in addition to normal 
outburst behavior. Superhumps, periodic signals which are presumed to 
arise from the prograde precession of an elliptically elongated accretion disk \citep{hellier:2001}, 
have previously been identified in the V1504 light curves and studied in detail 
by \citet{cannizzo:2012}.

A thorough analysis of the \emph{Kepler} lightcurve of V1504 Cygni 
suggests the presence of \textit{negative} superhumps 
(henceforth referred to by the shorter term infrahumps), as well. 
These are presumed to be associated with the nodal precession of a tilted
accretion disk. 

In this paper, we outline the process of identifying and extracting the period 
of these signals in Sec. \ref{sec:data-analysis}, and then show 
that they are consistent with those predicted by the period excess-deficit 
relation in Sec. \ref{sec:discussion}. We conclude with our main findings in 
Sec. \ref{sec:conclusion}.

\section{Data Analysis}
\label{sec:data-analysis}
\begin{figure}
\includegraphics[width=84mm]{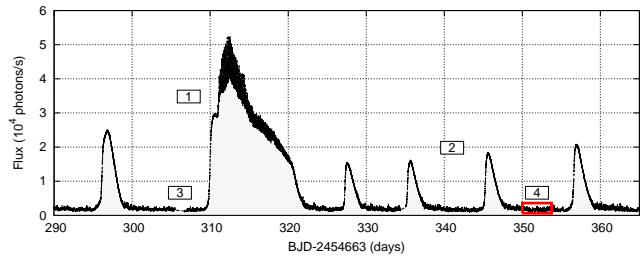}
\caption{A segment of the SAP light curve for V1504 Cygni spanning
from March 7, 2010 to June 15, 2010. Featured prominently in the
center is the 5th superoutburst observed by \textit{Kepler}. 
Notable features are (1) a superoutburst region with pronounced 
superhumps preceded by a shoulder with a profile matching a typical 
outburst, (2) outbursts that tend to gradually increase in magnitude, 
leading up to a superoutburst, (3) example data gap
that would be corrected before periodicity analysis 
is performed, and (4) a five day window between
350 and 355 days that is used to generate the plots in 
Figures \ref{fig:pds_plot} and \ref{fig:sine_fit} as indicated in
red.}
\label{fig:light_curve}
\end{figure}

The \textit{Kepler} data used in this study are the Simple Aperture 
Photometry (SAP) photon fluxes recorded in the \textit{Kepler} short 
cadence (SC) of approximately 1 minute. They are taken from a 550 day 
span starting on June 20, 2009 and lasting to December 22, 2010. 
A portion of the associated light curve is shown in 
Fig. \ref{fig:light_curve}.

Gaps in the data that correspond to monthly data downloads 
occur throughout the light curve, which if not 
treated properly can distort the temporal-resolution of the 
time-series or introduce anomalous structures in the associated 
power specta. These structures were ultimately eliminated through a 
correction process in which the gaps were filled with adjacent data, 
half from the left of each gap and half from the right. This has the benefit of reducing 
drastic oscillations in the high-frequency components of the power 
spectrum, enabling clearer recognition of peaks that correspond to 
physical frequencies of interest.

The improved clarity does not come without cost, however. When a longer 
gap (typically corresponding to scheduled data transfer or unexpected 
downtime) appears at the periphery of an outburst phase, part of the 
outburst is often copied into the gap, creating the appearance of a 
double-peak. This generates an artificial low-frequency contribution 
in the power spectrum for regions containing one of these gaps. This 
manifests in the power spectrum as 
intermittent vertical banding. This banding does not affect the identification 
or extraction of periodic signals in unaffected regions, and is 
therefore ignored for the purposes of this 
paper.\footnote{The \textit{Kepler} Science Office maintains a complete 
listing of known data artifacts, and regularly releases updates on any 
changes that present themselves, 
at http://archie.stsci.edu/kepler/data\_release.html.}

Once the SAP light curve was extracted from the \textit{Kepler} data 
files and corrected for relevant artifacts, residuals were obtained by 
subtraction of a boxcar-smoothed version of the curve. This was done 
largely to eliminate saturation in the low-frequency spectrum during 
subsequent analyses, in order to emphasize the presence of 
higher frequency periodic signals associated with the orbital 
dynamics of the system.

\begin{figure*}
\includegraphics[width=168mm]{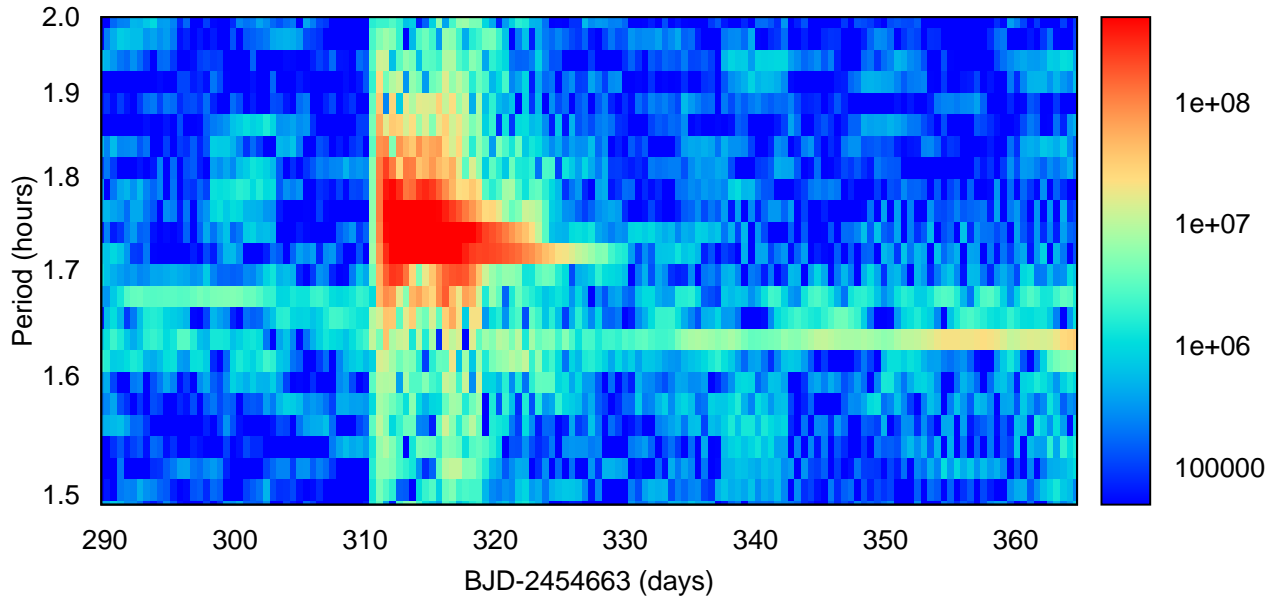}
\caption{
The power spectrum for the region of the light curve shown in 
Fig. \ref{fig:light_curve}, highest power output appears in 
red whereas the lowest appears in blue. The superoutburst that 
occurs near day 310 in Fig. \ref{fig:light_curve} 
is associated with a dramatic increase in 
the contribution from confirmed superhumps. Two additional, 
weaker signals are indicated by the horizontal banding observed 
at periods of $\sim$1.67 and $\sim$1.63 hours. 
These are confirmed to be the orbital, and infrahump periods, 
respectively.
}
\label{fig:heat_map}
\end{figure*}
\begin{figure}
\includegraphics[width=84mm]{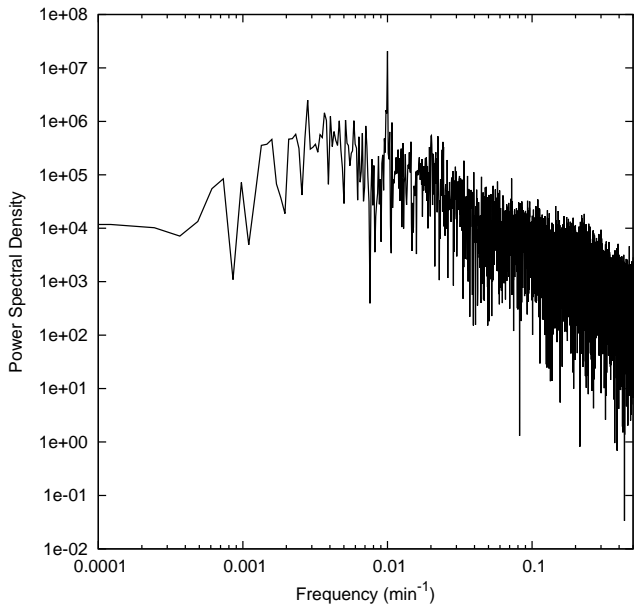}
\caption{
The power spectrum of the residuals for a 5 day window spanning 
from 350 to 355 (BJD-2454663) days, as indicated by the shaded 
box in Fig. \ref{fig:light_curve}. The peak at approximately
0.01 min$^{-1}$ corresponds to an infrahump period of $1.633\pm0.005$ hours.
}
\label{fig:pds_plot}
\end{figure}
\begin{figure}
\includegraphics[width=84mm]{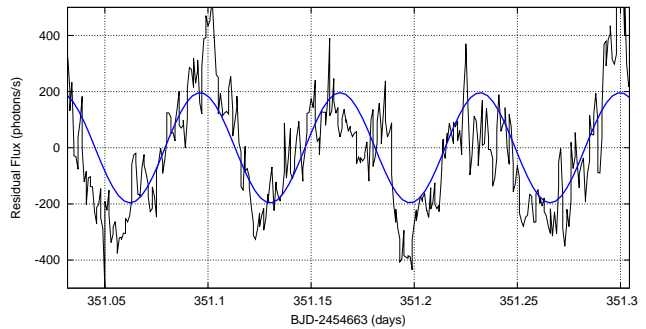}
\caption{
Example of the residual flux within the 350-355 (BJD-2454663) 
day window
highlighted in Fig. \ref{fig:light_curve}. The 
resultant sinusoidal fit is overlaid in blue. 
The fit result for the infrahump period is $1.629\pm0.024$ hours,
in good agreement with the values extracted from the 
power spectrum.
}
\label{fig:sine_fit}
\end{figure}

Following \citet{still:2010}, the power 
spectrum (Fig. \ref{fig:heat_map}) was obtained from the residuals by Short-time 
Fourier transform (STFT), which here aggregates the power spectra 
over a 120 hour window, stepped across the duration of the light 
curve in 12 hour increments. Though there are no permanent periodic 
structures, there are three distinct signals that appear 
with varying degrees of persistence and intensity. The 
periods associated with each signal were obtained by extracting 
peak frequency values from the power spectrum (see Fig. \ref{fig:pds_plot}). 
In addition, the extracted periods were confirmed by using a
Levenberg-Marquardt non-linear least squares algorithm to fit a sinsoidal 
function to selected quiescent regions of the light curve (see Fig. \ref{fig:sine_fit}). 

The most dramatic feature in Fig. \ref{fig:heat_map} is the 
presence of superhumps whose contributions dominate during the 
superoutburst phase. These occur when instabilities form during 
an outburst (hence the shoulder-like feature seen preceeding 
the superoutburst in Fig. \ref{fig:light_curve}), elongating the 
disk, which then begins to precess. These signals correspond to a 
period of $P_\mathrm{sh}=1.732\pm0.010$ hours, in excellent agreement
with the superhump period value measured by \citet{antonyuk:2005}.

There is also evidence for two additional periodic signals that 
arise in the latter half of the power spectrum. We identify the weaker of 
these two signals, which becomes apparent approximately 200 
days after the start of data acquisition, 
as the orbital period 
($P_\mathrm{o}=1.668\pm0.006$ hours), in good agreement with 
the value measured by \citet{thorstensen:1998}. 
Once manifested, this signal persists throughout the remainder of the data set. 

The second, stronger signal begins to contribute after 
approximately 250 days, and then intensifies, dominating 
the upper end of the frequency spectrum. It corresponds to a 
period of $P_\mathrm{ih}=1.633\pm0.005$ hours. There is strong 
evidence that this period corresponds to infrahumps, as will be 
discussed in Sec. \ref{sec:discussion}.

\begin{table}
\caption{Summary of salient results for V1504 Cyg.}
  \label{tab:histo}
  \begin{tabular}{@{}lcc}
    \hline
Parameter & Symbol & Value (units)\\
\hline
Orbital period & $P_\mathrm{o}$	&	$1.668\pm0.006$ (hours) \\
Superhump period & $P_\mathrm{sh}$	&	$1.732\pm0.010$ (hours)\\
Infrahump period & $P_\mathrm{ih}$	&	$1.633\pm0.005$ (hours)\\\\
Superhump period excess & $\epsilon_+$		&	$(3.84\pm0.70)$\% \\
Infrahump period deficit & $\epsilon_-$		&	$-(2.10\pm0.47)$\% \\\\
Mass ratio & $q$	& $0.168\pm0.044$\\
Secondary mass & $M_2$ & $0.128\pm0.050$ ($M_\odot$) \\
Roche lobe volume radius & $R_2/a $	& $0.246\pm0.022$\\
Apsidal precession period & $P_\mathrm{ap}$	&	$45.1\pm8.1$ (hours)\\
Nodal precession period & $P_\mathrm{no}$	&	$77.8\pm17.3$ (hours)\\
\hline
\end{tabular}
\medskip
\end{table}

\section{Discussion}
\label{sec:discussion}

It is useful to discuss the nature of a dwarf nova's orbital dynamics in terms of 
the so-called superhump period excess, defined as
\begin{equation}
\epsilon_+\equiv\frac{P_\mathrm{sh}-P_\mathrm{o}}{P_\mathrm{o}},
\label{eq:period-excess}
\end{equation}
which is a measure of fractional apsidal 
precession. \citet{patterson:1998} and \citet{murray:2000} 
showed that this value is dependent on the mass ratio 
of the system, which itself can be recast in terms of the orbital 
period. This not only provides a useful metric for quantifying 
whether or not an observed periodic signal is associated with 
superhumps, but also for obtaining the mass-ratio due to the readily 
accessible measurement of orbital period. Using the formulation 
presented by \citet{patterson:2005}, the superhump period excess 
is related to the mass ratio of 
the binary system by
\begin{equation}
\epsilon_+=0.18q+0.29q^2.
\label{eq:patterson-rel}
\end{equation}
With the assumed mass of a white dwarf ($M_1=0.76\pm0.22~M_\odot$) and the 
\citet{smith:1998} secondary mass-period relation 
\begin{equation}
M_2\approx(0.038\pm0.003)P_o^{(1.58\pm0.09)}~M_\odot,
\label{eq:smith-dhillon-rel}
\end{equation}
it is straightforward to calculate the mass ratio from an observed 
orbital period. Here, the measured superhump period excess for 
V1504 results in a mass ratio of $q=0.1677\pm0.0086$, 
well below the critical value that differentiates between 
\textit{SU UMa}- and \textit{U Gem}-type systems and is therefore required for 
the emergence of superhumps, confirming V1504's classification. 
Likewise, an estimate for the mass of the secondary can be placed 
at $0.1275\pm0.0375~M_\odot.$

\begin{figure} 
\includegraphics[width=84mm]{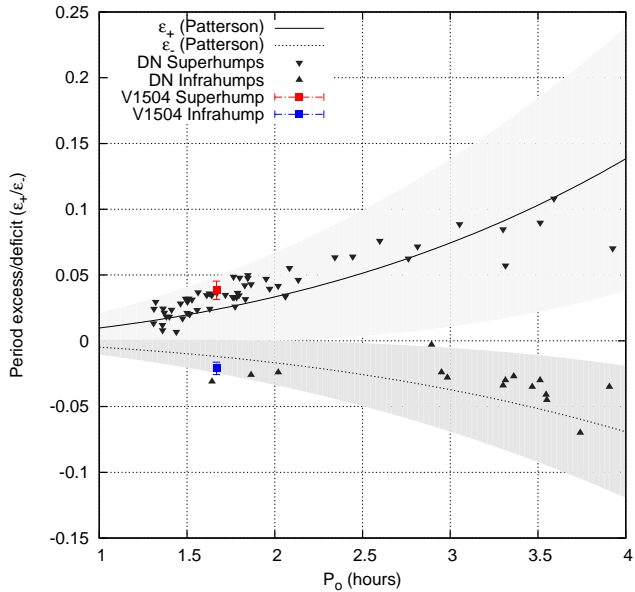} 
\caption{Theoretical dependence on orbital period for superhump 
period excess (solid line) and infrahump period deficit (dotted line) 
from \citet{patterson:2005}. The shaded regions associated with 
each line reflect the uncertainty in the values used in 
Eq.~\ref{eq:smith-dhillon-rel}. Plotted as grey triangles are the 
values for superhumps using data from \citet{patterson:1998} and 
infrahumps using a selection of data compiled by \citet{wood:2009} 
from numerous sources. The V1504 values for superhump excess (red square) 
and infrahump deficit (blue square) are plotted accordingly, and 
fall comfortably within the error regions.} 
\label{fig:period_excess} 
\end{figure}

The superhump period excess is also valuable for determining whether 
or not an observed periodic signal is associated with infrahumps. 
Defining the infrahump period deficit ($\epsilon_-$) 
in a manner identical to Eq.~\ref{eq:period-excess}, one obtains a 
value related to the fractional nodal precession (negative since the 
precession is retrograde). This has been empirically shown by 
\citet{patterson:1999} 
to be directly related to the 
superhump period excess by $\epsilon_+/\epsilon_-=-2.$ 
This relationship, along with Eq.~\ref{eq:patterson-rel} and 
Eq.~\ref{eq:smith-dhillon-rel}, is used to plot the theoretical $\epsilon_\pm$ 
values for varying orbital periods, along with data from several other 
systems that display super- and infrahumps in Fig. \ref{fig:period_excess}. 
The measured value for the third periodic signal present in the power 
spectrum resides comfortably within the error bars associated 
with $\epsilon_-$, suggesting strongly that it is caused by infrahumps. 
Though this feature is not permanent, it does persist for a 
significant duration after manifesting, raising questions as to its 
source. \citet{kato:2012} suggest that such infrahumps may be 
the result of weak excitations of tidal instabilities in the disk that 
fail to trigger a superoutburst, which may imply a common origin for 
superhump and infrahump excitations.

These results, along with additional calculated values for apsidal and nodal 
precession from formulae described in \citet{hellier:2001} and the Roche lobe volume radius 
from formulae described in \citet{smith:1998}, are summarized in Table \ref{tab:histo}.

\section{Conclusion}
\label{sec:conclusion}

By examining the \textit{Kepler} short cadence data for V1504 Cygni, we have 
identified three distinct periodic signals. Two of these are readily 
identifiable as the orbital period and the well-studied superhump period, 
each confirmed by independent measurements.

The presence of infrahumps in V1504 is inferred from the measurement 
of a third periodic signal in the power spectrum that corresponds 
to neither the established orbital period nor superhump period, 
and confirmed by comparison of the infrahump period deficit against 
empirical models. We have measured the infrahump period 
to be $1.633\pm0.005$ hours.

\section{ACKNOWLEDGEMENTS}

The GW astro group is grateful for the help provided by Martin Still 
in the initial acquisition of some of the CV data. 

\bibliography{bibliography.bib}

\begin{thebibliography}{99}






\bibitem[Antonyuk \& Pavlenko(2005)]{antonyuk:2005}
	Antonyuk, O.I., Pavlenko, E.P.,\ 2005,
	SU UMa-type dwarf nova V1504 Cygni during several supercycles,
	ASP Conference Series, 330

\bibitem[Cannizzo et al.(2011)]{cannizzo:2012}
	Cannizzo J.K., Smale A.P., Wood M.A., Still M.D., Howell S.B.,\ 2012,
	The Kepler light curves of V1504 Cygni and V344 Lyra: a study of the outburst properties,
	ApJ, 747, 117

\bibitem[Hellier(2001)]{hellier:2001} 
	Hellier C., \ 2001, 
	Cataclysmic Variable Stars, How and why they vary, 
	Praxis Publishing, Chichester, UK

\bibitem[Kato et al.(2012)]{kato:2012} 
	Kato T. et al,\  2012, 
	Survey of Period Variations of Superhumps in SU UMa-Type Dwarf Novae. III: The Third Year (2010-2011), 
	PASJ, 64, 21

\bibitem[Koch et al.(2010)]{koch:2010}
	Koch D.G. et al,\ 2010,
	Kepler Mission Design, Realized Photometric Performance, and Early Science,
	ApJ, 713, L79
	
\bibitem[Murray(2000)]{murray:2000}
	Murray J.R.,\ 2000,
	The precession of eccentric discs in close binaries,
	MNRAS, 314, L1

\bibitem[Patterson(1998)]{patterson:1998}
	Patterson J.,\ 1998,
	Late Evolution of Cataclysmic Variables,
	PASP, 110, 752

\bibitem[Patterson(1999)]{patterson:1999}
	Patterson J., 1999, in Mineshige S., Wheeler J.C., eds,
	in Disk Instabilities in Close Binary Systems,
	Kyoto: Universal Acad. Press, 61

\bibitem[Patterson(2005)]{patterson:2005}
	Patterson J. et al,\ 2005,
	Superhumps in Cataclysmic Binaries. XXV. $q_\mathrm{crit},~\epsilon(q)$, and Mass-Radius,
	PASP, 837, 117


\bibitem[Smith \& Dhillon(1998)]{smith:1998}
	Smith D.A., Dhillon V.S.,\ 1998,
	The secondary stars in cataclysmic variables and low-mass X-ray binaries,
	MNRAS, 301, 767

\bibitem[Still et al.(2010)]{still:2010}
	Still M., Howell S.B., Wood M.A., Cannizzo J.K., Smale A.P.,\ 2010,
	Quiescent superhumps detected in the dwarve nova V344 Lyra by Kepler,
	ApJ, 717, L113
	
\bibitem[Thorstensen \& Taylor(1998)]{thorstensen:1998}
	Thorstensen J.R., Taylor C.J.,\ 1998,
	Orbital Periods for the SU UMa-Type Dwarf Novae UV Persei, VY Aquarii, and V1504 Cygni,
	PASP, 109, 1359

\bibitem[Wood, Thomas \& Simpson(2009)]{wood:2009}
	Wood M.A., Thomas D.M., Simpson J.C.,\ 2009,
	SPH Simulations of Negative (Nodal) Superhumps: A Parametric Study,
	MNRAS, 398, 2110
	
\bibitem[Wood et al.(2011)]{wood:2011}
	Wood M.A., Still M.D., Howell S.B., Cannizzo J.K., Smale A.P.,\ 2010,
	V344 Lyra: A touchstone SU UMa cataclysmic variable in the kepler field,
	ApJ, 741, 105
	




















































































































\end{thebibliography}

\end{document}